# Evaluation of Consumer Behavior Regarding Food Delivery Applications in India

**Sarvesh Jadhav**
**MBA - MARKETING**
Alliance University, Bangalore- INDIA

**Dr. Ray Titus**
Pro Vice-Chancellor (Incubation and Innovation )
Dean, ASOB
Alliance University , Bangalore – INDIA

**Dr.Tina Babu**
Associate Professor
Alliance University,Bangalore – INDIA

**Dr.R.Chinnaiyan**
Professor & Associate Director
Alliance University , Bangalore – INDIA

**Abstract.** This paper explores consumer behavior towards food delivery apps, focusing on attributes like restaurant variety, food packaging quality, and application design and user interface. The study reveals that a diverse range of restaurants positively influences consumer satisfaction, leading to increased app usage. Conversely, the quality of food packaging does not significantly impact overall satisfaction. However, the study underscores the crucial role of application design and user interface in shaping consumer behavior. Both factors significantly influence overall satisfaction, with user-friendly interfaces attracting more users and promoting frequent app usage. The findings emphasize the importance of businesses addressing these attributes to enhance customer satisfaction, boost app engagement, and foster long-term customer loyalty. Understanding and catering to consumer preferences in these areas can contribute to the success of food delivery apps in the competitive market.

## 1 Introduction

In recent times, the surge in popularity of food delivery apps has led to diverse customer behaviors influenced by factors such as user interface, application design, restaurant variety, and packaging quality. The dynamics of consumer engagement on these platforms are intricately linked to the interplay of these elements. Specifically, the user experience, shaped by the user interface and application design, significantly impacts consumer behavior. An intuitive, visually appealing, and well-functioning app tends to retain users, while a poorly designed one may lead to dissatisfaction and potential migration to competing apps [4].

Moreover, the variety of restaurants available on these platforms plays a crucial role in shaping customer behavior. The more extensive the selection, encompassing diverse cuisines and price ranges, the more likely customers are to use the app regularly. This variety caters to different preferences, with some customers opting for upscale dining on special occasions and others seeking more affordable options for everyday meals [6].

Additionally, the quality of packaging emerges as a factor influencing consumer behavior. Delivery experiences are positively impacted when food arrives in premium, durable packaging, maintaining its freshness and temperature. Conversely, subpar packaging that compromises the quality of delivered food can result in customer dissatisfaction, prompting them to explore alternative options [5].

Building on previous research, a study delved into the intricate connections among these factors that shape consumer utilization of food delivery applica- tions. Using an expanded flow theory model, the research explored consumers' ex- periences in ordering meals through mobile apps. The findings, analyzed through structural equation modeling, highlighted the substantial impact of consumer experience, both in web and digital domains, on purchasing intention [7].

## 2   Literature Survey

The objective of this study is to explore the risk and benefit perceptions associated with the utilization of online food delivery services (OFDAs) by Indian consumers, as well as the factors influencing their selection behavior. A survey of 337 OFDA users employing exploratory factor analysis identified 5 risk and 2 benefit factors, which were subsequently tested in a structural model with 31 constructs. The results revealed that various factors, encompassing consumers' general attitudes, habits, perceived risks, and rewards, influence their behavior and selection of OFDAs. Moreover, the study found that consumers' attitudes towards OFDA usage are subject to change based on their perceptions of reduced risks or increased benefits [8].

Another research project, focused on international students at the Manipal Academy of Higher Education, examined the impact of mobile app design elements on decision-making in the online food ordering and delivery sector. Utilizing Smart PLS-3 SEM analysis, the study identified several design features significantly influencing students' behavior, including security of personal information, accessible payment gateways, real-time order tracking, customer service accessibility, trending information pop-ups, easy navigation, search functionality, and flexible delivery services [1].

In a separate study, the factors influencing customer satisfaction with food delivery apps in India were explored [10]. The study highlighted meal quality, delivery time, and app user interface as key determinants of customer satisfaction, emphasizing the crucial role of these factors in users' overall satisfaction with meal delivery applications [9].

Furthermore, an investigation into the technological and non-technology components of OFD services explored their impact on attitudes and behavioral intentions. The study examined the relationship between brand image, brand love, and various factors such as personality, social impact, service quality, and app design. Notably, app design was identified as a significant factor affecting brand image, while service excellence, personality characteristics, and social influence had a comparatively lesser impact on brand image growth. The study also identified that brand image fully mediated the relationship between app design, service quality metrics, and brand love [2].

Lastly, a study focusing on the psychological and technological aspects influencing consumers' decisions regarding meal delivery apps investigated perceived utility, incentives, information, customer relationship management, and order management system as positive influencers of customer conversions. In contrast, perceived cost and aesthetic appeal were found to have minimal impact [3].

In summary, the literature survey provides a comprehensive overview of the multifaceted dynamics in the consumer landscape of online food delivery services in India. The studies collectively contribute insights into the diverse factors shaping consumer behavior, preferences, and satisfaction in this rapidly evolving sector.

## 3   Research Methodology

The identified research gap suggests a potential correlation between the quality of food packaging and consumer behavior, potentially influencing users to switch  to alternative applications or discontinue app usage. Additionally, a complex  user interface may lead to customer impatience, impacting immediate or future purchases. The following are the research objectives identified

- Examine Consumer Behavior: Investigate the intricacies of consumer behavior concerning food delivery apps.
- Assess Application Design Impact: Evaluate how application design influences consumer behavior in the context of food delivery apps.
- Analyze Restaurant Variety Impact: Study the effects of a wide variety of restaurants on consumer behavior in relation to food delivery apps.
- Evaluate UI/UX Impact: Examine the impact of User Interface/User Experience (UI/UX) on consumer behavior towards the use of food delivery apps.
- Investigate Food Packaging Impact: Explore the influence of food packaging on consumer behavior concerning the utilization of food delivery apps.

### 3.1   Hypothesis

1. Effect of a variety of restaurants
    - $H_0$: Impact of wide variety of restaurants does affect consumer behaviour towards food delivery apps.

- $H_1$: Impact of wide variety of restaurants does not affect consumer behaviour towards food delivery apps.
2. Effect of quality of packaging
3. - $H_0$: Impact of quality of packaging does affect consumer behaviour towards food delivery apps.
   - $H_1$: Impact of quality of packaging does not affect consumer behaviour towards food delivery apps.
4. Effect of application design
   - $H_0$: Impact of application design does affect consumer behaviour towards food delivery apps.
   - $H_1$: Impact of application design does not affect consumer behaviour towards food delivery apps.
5. Effect of user interface
   - $H_0$: Impact of user interface does affect consumer behaviour towards food delivery apps.
   - $H_1$: Impact of user interface does not affect consumer behaviour towards food delivery apps.

### 3.2 Data Collection Techniques

The study will be focused on Indian food delivery apps and how consumers react to these applications based on variables such as application design, user interface, a wide variety of restaurants, and quality of packaging. For data collection, two types of data would be collected,

1. Primary Research: To gather data for primary research, two surveys have been carried out. The data obtained through these two surveys will be an- alyzed by JAMOVI to derive conclusions after the sample size has been reached and the data has been collected.
   Location for study: Pan India, Sample size: 229
2. Secondary Research: The secondary data has been collected from articles, newspapers, journals, and the internet.

## 4 Data Analysis and Interpretation

### 4.1 Based on Survey 1

The survey was conducted on Google Forms. The total number of responses collected by the survey was 229 responses.

While the respondents' age group exhibits diversity, a majority (190) are aged 20-30, indicating a focus on this demographic (Figure 1(a)). In terms of location (Figure 1(b)), respondents are predominantly from metro cities (e.g., Mumbai, Delhi), followed by Tier 1 non-metros (e.g., Pune), rural areas, and Tier 2 cities (e.g., Bareilly). This suggests a comparative analysis of purchasing behavior between metro cities and Tier 1 non-metros.

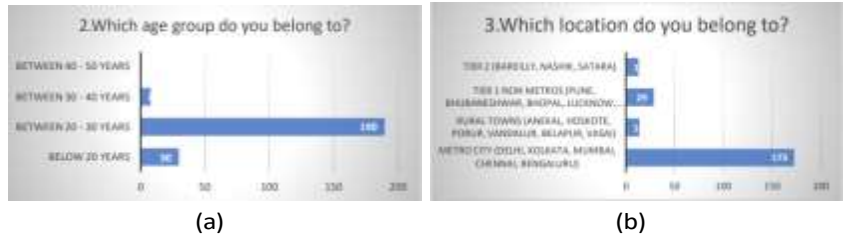

**Fig. 1.** Analysis based on (a) Age (b) Location.

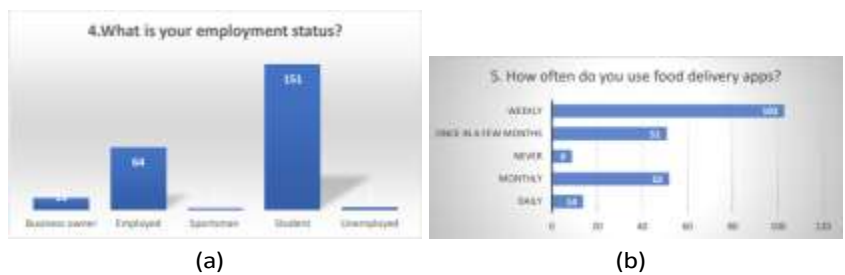

**Fig. 2.** Analysis based on (a) Employment Status (b) Usage of delivery apps.

Figure 2(a) indicates that respondents are primarily students (151), followed by employed individuals (64), and others such as business owners (11), unem- ployed, and sports professionals. This comparison focuses on the purchasing be- havior of students, likely from Gen Z, and the employed segment spanning ages 20-30 to 51 and above. Regarding the frequency of product purchases over time (Figure 2(b)), the majority prefer weekly purchases (103), followed by monthly (52), once in a few months (51), daily (14), and never (9).

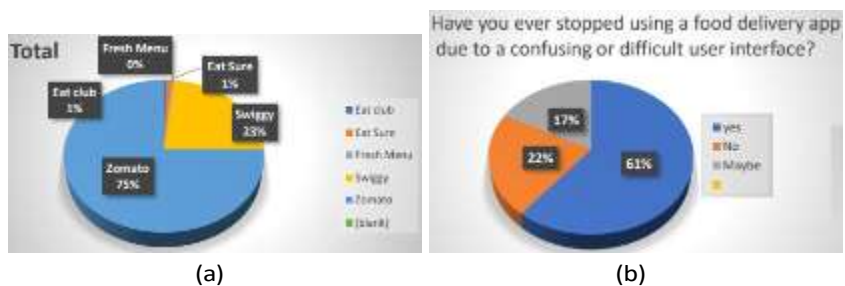

**Fig. 3.** Analysis based on (a) App Usage (b) Stopping the usage of apps.

Figure 3(a) displays the purchase frequency from listed apps, with Zomato being the most preferred (165), followed by Swiggy (51), Eat Sure (2), Eat Club, and Fresh Menu (1). Figure 3(b) indicates that a majority of respondents (61%) stopped using food delivery apps due to confusion or a difficult user interface, emphasizing the importance of a user-friendly interface in building brand value and expanding user base.

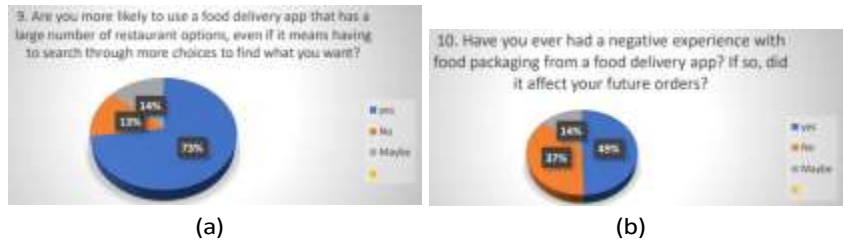

(a)                                    (b)

**Fig. 4.** Analysis based on (a) Restaurant options (b) Negative experience.

Figure 4(a) indicates that a majority of respondents (73%) are inclined to use a food delivery app with numerous restaurant options, even if it means navigating through more choices. In Figure 4(b), the majority of respondents (49%) affirmed that a negative experience with food packaging from a food delivery app does impact their future orders, emphasizing the significance of packaging in customer.

## 4.2   Based on Survey 2

The survey was conducted on Google Forms. The following are the results of the survey

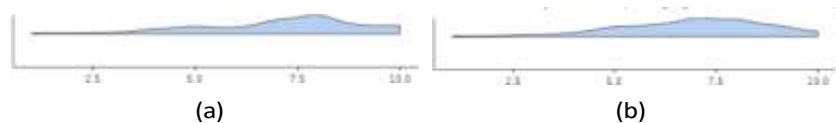

(a)                                    (b)

**Fig. 5.** Analysis based on (a) Satisfaction on Restaurant options (b) Packaging satisfaction.

Figure 5(a) depicts the satisfaction levels of respondents regarding the diverse restaurant options on food delivery apps, with the majority rating between 6 and 9. In Figure 5(b), respondents' satisfaction with the packaging of delivered food is shown, with the majority giving ratings between 5 and 9.

Figure 6(a) illustrates respondents' satisfaction with the application design of food delivery apps, with many falling within the rating range of 5 to 10. In

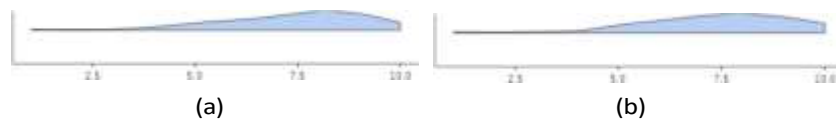

**Fig. 6.** Analysis based on (a) App Design (b) Overall satisfaction of apps.

Figure 6(b), the overall satisfaction level of respondents with the usage of food delivery apps is presented, with most ratings falling within the range of 6 to 10.

## 5    Findings

From Table 1 and Table 2, the following observations have considered.

- **Pair 1:** Overall satisfaction on usage of food delivery apps and choosing from wide variety of restaurants listed on food delivery apps.
  The p-value is lower than the significance level i.e. 0.001 in this case. It suggests that the observed data is statistically significant.In this situation, it can be concluded that choosing from wide variety of restaurants will affect the overall satisfaction as the p-value is below the significance level (0.05).
  • H0: The impact of a wide variety of restaurants does affects consumer behaviour towards food delivery apps.
  • H1: Impact of wide variety of restaurants does not affect consumer behaviour towards food delivery apps.
  Thus, it will accept the null hypothesis and reject the alternate hypothesis.
- **Pair 2:** Overall satisfaction on usage of food delivery apps and packaging of food delivered by food delivery apps.
  The p-value is higher than the significance level i.e. 0.366 in this case. It sug- gests that the observed data is statistically not significant.In this situation, it can be concluded that quality of food packaging will not affect the overall satisfaction as the p-value is above the significance level (0.05)
  • H0: Impact of quality of packaging does affect consumer behaviour to- wards food delivery apps.
  • H1: Impact of quality of packaging does not affect consumer behaviour towards food delivery apps.
  Thus, it will accept the alternate hypothesis and reject the null hypothesis.
- **Pair 3:** Overall satisfaction on usage of food delivery apps and application design of food delivery apps.
  The p-value is lower than the significance level i.e., ¡ 0.001 in this case. It suggests that the observed data is statistically significant.In this situation, it can be concluded that the application design of food delivery apps will affect the overall satisfaction as the p-value is below the significance level (0.05).
  • H0: Impact of application design does affect consumer behaviour towards food delivery apps.

- • H1: Impact of application design does not affect consumer behaviour towards food delivery apps.

  Thus, it will accept the null hypothesis and reject the alternate hypothesis.
- — **Pair 4:** Overall satisfaction on usage of food delivery apps and user interface of food delivery apps.

  The p-value is lower than the significance level i.e., ¡ 0.001 in this case. It suggests that the observed data is statistically significant. In this situation, it can be concluded that the user interface of food delivery apps will affect the overall satisfaction as the p-value is below the significance level (0.05).
  - • H0: Impact of user interface does affect consumer behaviour towards food delivery apps.
  - • H1: Impact of user interface does not affect consumer behaviour towards food delivery apps.

  Thus, it will accept the null hypothesis and reject the alternate hypothesis.

**Table 1.** Linear Regression.

| Model Fit Measures | | |
|---|---|---|
| Model | R | $R^2$ |
| 1 | 0.795 | 0.632 |

**Table 2.** Model Coefficients - Overall satisfaction on usage of food delivery apps.

| Predictor | Estimate | SE | t | p |
|---|---|---|---|---|
| Intercept | 0.5972 | 0.4154 | 1.438 | 0.152 |
| How satisfied are you with choosing from the wide variety of restaurants listed on food delivery apps? | 0.1925 | 0.0586 | 3.287 | 0.001 |
| How satisfied are you with the packaging of the food delivered by food delivery apps? | 0.0526 | 0.0580 | 0.907 | 0.366 |
| How satisfied are you with the Application design of food delivery apps? | 0.2930 | 0.0753 | 3.889 | ¡.001 |
| How satisfied are you with the User Interface of food delivery apps? | 0.3914 | 0.0745 | 5.257 | ¡.001 |

**Table 3.** Linear Regression.

| Model Fit Measures | | |
|---|---|---|
| Model | R | $R^2$ |
| 1 | 0.0566 | 0.00320 |

**Table 4.** Model Coefficients 2 - Overall satisfaction on usage of food delivery apps.

| Predictor | Estimate | SE | t | p |
|---|---|---|---|---|
| Intercept | 7.515 | 0.225 | 33.367 | ¡ 0.001 |
| Gender: Male / Female | -0.216 | 0.285 | -0.761 | 0.448 |

From Table 3 and Table 4, the p-value = 0.448 which is above the significance level. Thus, it suggests that the observed data is statistically not significant. In this situation, it can be concluded that the gender of respondents will not affect the level of overall satisfaction as the p-value is above the significance level (0.05).

## 6 Conclusion

This research has delved into consumer behavior regarding food delivery applications, assessing the influence of key attributes such as restaurant variety, food packaging quality, and application design and user interface. Analyzing existing literature, several crucial findings have emerged. Firstly, the study establishes a positive impact of a diverse array of restaurants on consumer behavior. The statistical analysis, with a p-value of 0.001 below the significance level, indicates a substantial influence. Consumers are inclined to engage with apps offering diverse dining options, enhancing convenience, and fostering increased usage and satisfaction. Secondly, the research concludes that the quality of food packaging does not significantly affect consumer behavior, as evidenced by a p-value of 0.366, surpassing the significance level. This implies that packaging quality does not substantially impact overall satisfaction.

Moreover, the study highlights the substantial impact of application design and user interface on consumer behavior and satisfaction. Both factors exhibit statistically significant p-values (¡ 0.001), emphasizing their influence. A user-friendly and visually appealing interface contributes to heightened user engagement, satisfaction, and continued app usage. Conversely, poorly designed interfaces may deter users, leading to reduced engagement and potential customer loss.

In summary, this dissertation underscores the critical role of understanding consumer behavior and the impact of specific attributes in the realm of food delivery apps. Addressing factors like restaurant variety, food packaging quality, and effective app design can enhance customer satisfaction, drive increased app engagement, and foster enduring customer loyalty.